\documentclass[aps,prb,twocolumn,showpacs,superscriptaddress,groupedaddress,nofootinbib]{revtex4-1}

\pdfoutput=1
\usepackage{graphicx}
\usepackage[utf8]{inputenc}
\usepackage{amsmath}
\usepackage{flushend}
\usepackage{dcolumn}
\usepackage{bm}
\usepackage{balance}

\usepackage{color,ulem}
\usepackage[english]{babel}

\usepackage[utf8]{inputenc}

\begin{document}

\preprint{IFT-UAM/CSIC-20-3}

\title{Effective Theory of Superconductivity \\in Strongly-Coupled Amorphous Materials}

\author{Matteo Baggioli}%
 \email{matteo.baggioli@uam.es}
\affiliation{Instituto de Fisica Teorica UAM/CSIC, c/Nicolas Cabrera 13-15,
Universidad Autonoma de Madrid, Cantoblanco, 28049 Madrid, Spain.
}%
\author{Chandan Setty}%
 \email{settychandan@gmail.com }
\affiliation{Department of Physics, University of Florida, Gainesville, Florida, USA}%

\author{Alessio Zaccone}%
 \email{alessio.zaccone@unimi.it }
\affiliation{
Department of Physics "A. Pontremoli", University of Milan, via Celoria 16, 20133 Milan, Italy.\\
Department of Chemical Engineering and Biotechnology,
University of Cambridge, Philippa Fawcett Drive, CB30AS Cambridge, U.K.\\
Cavendish Laboratory, University of Cambridge, JJ Thomson
Avenue, CB30HE Cambridge, U.K.
}

\begin{abstract}
A theory of phonon-mediated superconductivity in strong-coupling amorphous materials is developed based on an effective description of structural disorder and its effect on the vibrational spectrum. The theory accounts for the diffusive-like transport of vibrational excitations due to disorder-induced scattering within the Eliashberg theory of strong-coupling superconductivity. The theory provides a good analytical description of the Eliashberg function $\alpha^{2}F(\omega)$ in comparison with experiments, and allows one to disentangle the effects of transverse and longitudinal excitations on the Eliashberg function. In particular, it shows that the transverse excitations play a crucial role in driving an increase or excess in the Eliashberg function at low energy, which is related to the boson peak phenomenon in vibrational spectra of glasses. This low-energy excess, on one hand drives an enhancement of the electron-phonon coupling but at the same time reduces the characteristic energy scale $\omega_{log}$ in the Allen-Dynes formula. As a consequence, the non-monotonicity of $T_{c}$ as a function of alloying (disorder) in $\text{Pb}$-based systems can be rationalized.
The case of $\text{Al}$-based systems, where disorder increases $T_{c}$ from the start, is also analyzed. General material-design principles for enhancing $T_{c}$ in amorphous superconductors are presented.
\end{abstract}

\maketitle

\section{Introduction}
Since its discovery in vapor-deposited amorphous films by Buckel and Hilsch~\cite{Buckel1954,Buckel1956}, superconductivity in amorphous (structurally disordered) materials has attracted much interest both from the experimental and from the theoretical/computational point of view. 
Since many different forms of "disorder" are possible in condensed matter systems (structural, electronic, magnetic, substitutional etc) we should focus here on structural disorder, i.e. disorder in the spatial arrangement of atoms such as the one encountered in amorphous metal alloys (also known as metallic glasses).\\
Typically, superconductivity in amorphous materials is strongly-coupled, with the electron-phonon coupling parameter $\lambda >1$ and the superconducting gap much larger than the BCS prediction, $\Delta \gg 1.76\, k_B\,T_c \equiv \Delta_{BCS}$. Hence, it should be described using the Eliashberg theory of strong-coupling phonon-mediated superconductivity. \color{black} A large value of the  ratio of the superconducting  gap to the critical temperature in phonon-mediated superconductivity  could be also due to phase  fluctuation  effects (see~\cite{Emery1995} and references therein). For clarity, the latter mechanism will not be discussed in this manuscript.\color{black}\\

Early work established that structural atomic disorder may promote an increase of $\lambda$ upon going from the crystalline to the amorphous state. \color{black}Some elements which are poor superconductors in the crystalline phase, such as Ga and Bi, or they do not superconduct at all such as Be~\cite{Buzea_2004}, become superconducting in the amorphous state~\cite{Bergmann1976,Curzon_1969}.\color{black} This can be explained with the fact that these elements do not have a dense packed structure in the crystalline phase, but rather an open lattice structure, hence disorder promotes superconductivity by effectively densifying the structure~\cite{Bergmann1976}. Other elements such as Pb are good superconductors in the crystalline phase thanks to their dense packed structure, and, in this case, the introduction of disorder lowers the $T_{c}$~\cite{Knorr1971,Bergmann1976}. 

In general, good crystalline superconductors are insensitive to dilute (non-magnetic) impurities, as predicted early on by Anderson~\cite{Anderson}. A sizable effect of disorder becomes apparent when the lattice structure becomes amorphous. Typically, structural disorder leads to a monotonic decrease of $T_{c}$, even when $\lambda$ increases with disorder, due to the faster decrease of the characteristic energy scale with disorder. In some cases even a superconductor-insulator transition (SIT) is observed~\cite{Ovadyahu1992}. The instances where disorder actually increases $T_{c}$ in amorphous superconductors are relatively rare and limited to the cases discussed above (where lattice structure is "open" in the crystal phase), and include situations where disorder is promoted by alloying or granular superconductors~\cite{Garland,Abeles,Ovadyahu}. 
Furthermore, understanding the effect of structural disorder on superconductivity is crucial also for high-$T_{c}$ materials such as cuprates, which are affected by substitutional disorder being out of stoichiometry~\cite{Klein2001,Leroux2019}.


A reason for this apparently scattered evidence in superconductors may also be due to the lack of theoretical frameworks able to disentangle and describe, in a reductionist way, the effect of structural disorder on the strong-coupling superconductivity. This delay of theory is imputable to the difficulty of providing an effective description of structural disorder on the vibrational spectrum of amorphous solids, where a long-standing issue is represented by the the so-called boson peak or excess of soft vibrational modes which  shows up in the VDOS upon normalizing it by the Debye law $\omega^{2}$~\cite{Schirmacher,Milkus}, and for which a deeper understanding has emerged only recently as follows.

Recent work has highlighted that phonons start to behave as non-ballistic excitations already at rather low wavevectors $k$ due to disorder-induced scattering. It has been established, with the help of numerical simulations, that the largest part of vibrational density of states (VDOS) of amorphous materials is made of diffusive-like excitations, i.e. excitations which propagate with a diffusive law due to intense scattering promoted by disorder. These excitations are sometimes referred to as "diffusons"~\cite{Allen_diffusons} and also play an important role in controlling the (low) thermal conductivity of amorphous materials~\cite{Allen1993}.
The crossover from ballistic phonons at low $k$ to diffusons at higher $k$ is of the Ioffe-Regel type, i.e. the wavelength of the phonon becomes comparable to the mean free path at some value of $k$\,\,\,\cite{Parshin1,Parshin2}. This crossover is also responsible for a pronounced peak in the VDOS, often referred to as the "boson peak", since in Raman scattering experiments it exhibits the same $T$-dependence as the Bose function, thus pointing at the "harmonic" origin of this effect, consistent with the diffusive-like nature of the excitations being due to elastic scattering events~\cite{Schirmacher}.

Recently, a theoretical model has been proposed which provides a unified description of the VDOS and phonon spectrum of amorphous materials based on an effective-field description of phonons/diffusons~\cite{baggioli2019unified}. In this work we adopt this effective description of vibrations in disordered systems as the starting point to evaluate the electron-phonon coupling semi-analytically, and to analyze the effect of structural disorder on superconductivity in amorphous materials at a deeper level. 
The theoretical model is able to provide successful fittings of the Eliashberg function for various amorphous materials, with very few physical parameters such as speeds of sound and linewidths (the latter related to the diffusivity $D$ of the excitations). The model is able to predict a number of results, including the non-monotonic behaviour of $\lambda$ as a function of disorder (depending on the value of speed of sound) that was already predicted at the level of BCS  theory~\cite{Chandan}, and to provide qualitative estimates of $T_{c}$ which feature highly non-trivial trends as a function of disorder as discussed below in the application to Pb- and Al-based materials.

\section{Theoretical Framework}
Our starting point is an equation of motion for the elastic displacement field $\mathbf{u}$ in an isotropic solid (i.e. such that longitudinal and transverse components can be decoupled). Aside from the standard elastodynamic part coming from linear elasticity theory, the equation contains a term describing momentum diffusion. In this way, the time evolution of the displacement field can cover both the ballistic (\textit{propagon}) and the diffusive (\textit{diffuson}) regimes of vibrational excitations, with the latter regime becoming comparatively more prominent upon increasing $k$, close and beyond the Ioffe-Regel crossover. The full dynamics is then described by the simple differential equation:
\begin{equation}
\frac{\partial^2 u_{\lambda}}{\partial t^2}=\, v_{\lambda}^2 \bigtriangleup u_{\lambda}+ D_{\lambda}\frac{\partial\bigtriangleup u_{\lambda}}{\partial t},\label{unob}
\end{equation}
where $v_\lambda$ and $D_\lambda$ are the (dressed) speed of propagation and the diffusion constant of the $\lambda$ branch, respectively. The subscript $\lambda$ refers to either longitudinal $\lambda=L$ or transverse $\lambda=T$ displacement field.\\
 In the following we shall consider both $L$ as well as $T$  excitations, because in amorphous solids there is no momentum conservation in electron-phonon scattering, hence in the derivation of the Eliashberg equations also the $T$ phonons contribute to the pairing, as the dot products do not vanish, differently from what happens in ordered superconductors where momentum conservation and Umklapp conditions lead to the cancellation of the transverse phonon contributions \cite{Lee,Gorkov}.

Upon Fourier transforming the displacement field $\mathbf{u}$, one readily obtains the associated Green function or propagator,
\begin{equation}
    G_\lambda(\omega,k)\,=\,\frac{1}{\omega^2\,-\,\Omega_{\lambda}^{2}(k)\,+\,i\,\omega\,\Gamma_{\lambda}(k)} \label{eq2}
\end{equation}
where the propagating term is given by $\Omega_{\lambda}^{2}= v_{\lambda}^2\,k^2\,$. The diffusive damping due to harmonic disorder-induced scattering, following from Eq.\eqref{unob}, is given by $\Gamma_{\lambda}(k)=D_\lambda\,k^2$. This diffusive form of the damping is supported by several simulation studies \cite{Tanaka2008,Parshin1,Parshin2,Schirmacher} over a broad range of $k$. \color{black} Following Ref.\cite{Allen_diffusons}, and comparing with our Eq. \eqref{eq2}, we can define three separate regimes and corresponding excitations. (I) The \textit{propagons}. This is the limit in which dissipation is not dominant, $v k \gg D k^2$ and the excitations are still well defined quasiparticles undergoing a ballistic motion. (II) The \textit{diffusons}. This regime appears beyond the Ioffe-Regel limit $v k \sim D k^2$ at which the dissipative term becomes dominant. At this point, there are not well-defined quasiparticles anymore and the dynamics is totally incoherent, collective and diffusive. In this range, it does not make sense to think of ballistic concepts such as the mean free path of propagation, simply because there are no propagating particles. (III) The \textit{locons}. This is the extreme limit, usually relevant close to the edge of mobility near the Debye frequency, in which the modes are completely (Anderson) localized. Not only the modes do not propagate but also their diffusion constant vanishes. \color{black} The \textit{diffusons} are also possibly related to the random-matrix structure of the eigenvalue spectrum of the dynamical (Hessian) matrix, as suggested in Refs.~\cite{BaggioliPRR}.

Let us remark that the liquid-like diffusive dynamics of the \textit{diffusons} is governed by the Einstein law \cite{doi:10.1002/andp.19053220806}:
    $D\,\sim\,\langle x^2 \rangle/\tau$,
where $\tau$ is the  characteristic relaxation time between two successive jumps, which, in this regime, decreases upon making the amount of disorder larger.

The above propagator leads to a dynamic structure factor (spectral function) for inelastic processes
\begin{equation}
    \mathcal{B}_{\lambda}(\omega,k)\,=\,-\frac{1}{\pi}\text{Im} G_\lambda(k, \omega+i\delta)
\end{equation}
where $\delta = \omega \,\Gamma$.
Using the propagator of Eq.\eqref{eq2}, this leads to a Lorentzian function for the inelastic dynamic structure factor
\begin{equation}
    \mathcal{B}(\omega,k)\,=\,\frac{\omega\,\Gamma(k)}{\pi[\left(\omega^2\,-\,\Omega^2(k)\right)^2\,+\,\omega^2\,\Gamma^2(k)]}\label{eq4}
\end{equation}
which has been used many times in the past to describe inelastic X-ray scattering data of vibrational excitations of amorphous materials~\cite{Benassi1996,Pilla2000,Baldi2008,Baldi2011}, as well as MD simulations data of disordered solids~\cite{Tanaka2008}.

The Eliashberg electron-phonon spectral function can be written as~\cite{2001cond.mat..6143M}:
\begin{equation}
    \alpha^2\,F(\Vec{k},\Vec{k}',\omega)\,\equiv\,\mathcal{N}(\mu)\,|g_{\Vec{k},\Vec{k}'}|^2\,\mathcal{B}(\Vec{k}-\Vec{k}',\omega)
\end{equation}
where $\mathcal{N}(\mu)$ is the electronic density of states (EDOS) at the chemical potential $\mu$ and $g_{\Vec{k},\Vec{k}'}$ the electron-boson matrix element. Following Ref.\cite{2001cond.mat..6143M}, the Fermi-surface-averaged spectral function is:
\begin{equation}
    \alpha^2\,F(\omega)\,=\,\frac{1}{\mathcal{N}(\mu)^2}\,\sum_{\Vec{k},\Vec{k}'}\,\alpha^2\,F(\Vec{k},\Vec{k}',\omega)\,\delta(\epsilon_{\Vec{k}}\,-\,\mu)\,\delta(\epsilon_{\Vec{k}'}\,-\,\mu)\, \label{ee}
\end{equation}
After assuming constant matrix elements $g_{\Vec{k},\Vec{k}'}\equiv \mathrm{g}$, Eq.\eqref{ee} for a one-phonon branch can be re-written as:
\begin{equation}
   \alpha^2\,F(\omega)\,=\,\frac{\mathrm{g}^2}{\mathcal{N}(\mu)}\,\sum_{\Vec{k},{\Vec{k}}'}\,\mathcal{B}(\Vec{k}-{\Vec{k}}',\omega)\,\delta(\Vec{k}^2\,-\,\mu)\,\delta({{\Vec{k'}}}^2\,-\,\mu)\,
\end{equation}
in which we took a quadratic electronic band $\epsilon_{\Vec{k}}=\Vec{k}^2$. We compute the previous sum by converting it to a two dimensional integral using $
    \sum_{\Vec{k}}=\,\frac{V_2}{(2\pi)^2}\,\int \,k\, dk\, d\phi_k $
with $k \in [0,\infty]$ and $\phi_k\in [0,2\pi]$. The last step is to notice that $\mathcal{B}(\Vec{k}-\Vec{k}',\omega)$ is only a function of $( \Vec{k}\,-\Vec{k}')^2$ and that such distance in polar coordinates can be expressed as $
   ( \Vec{k}\,-\Vec{k}')^2\,=\,k^2\,+\,{k'}^2\,-\,2\,k\,k'\,\cos(\phi_k\,-\,\phi_{k'})$.\\
All in all, we can perform immediately the integral over the moduli using the definition of delta functions and we obtain the final result:
\begin{align}
    &\alpha^2\,F(\omega)\,=\,\frac{\mathrm{g}^2}{4\,(2\pi)^4\,\mathrm{N}}\,\int \sum_{\lambda} \mathcal{B}_{\lambda}(X^2,\omega)\,d\phi_k\,d\phi_{k'} \label{aa}\\
    &\mathcal{B}(X^2,\omega)\,=\, \frac{\omega\,D\,X^2}{\pi\left(\omega^2\,-\,v^2\,X^2\right)^2\,+\,\omega^2\,D^2\,X^4}\label{eq9}\\
    &  X^2\,\equiv\,2\,\mu\,\left(1\,-\,\cos(\phi_k\,-\,\phi_{k'})\right) \label{eq10}
\end{align}
where the sum runs over phonon branches $\lambda$, and we used the fact that the electronic density of states in two dimensions is constant, $\mathcal{N}(\mu)=\mathrm{N}$. This final integral in Eq.\eqref{aa} can be performed numerically.\\

At this point, we can use the standard definition for the electron-phonon mass enhancement parameter:
\begin{equation}
    \lambda(v,D)\,=\,2\,\int_0^\infty\, \frac{\alpha^2\,F(\omega)}{\omega}\,d\omega\label{eq11}
\end{equation}
determining the effective (dimensionless) strength of the electron-phonon interactions.
In order to estimate the critical temperature $T_{c}$, we use the Allen-Dynes formula ~\cite{PhysRevB.12.905} given by:
\begin{equation}
      T_c\,=\,\frac{f_1\,f_2\,\omega_{log}}{1.2}\,\exp\left(-\frac{1.04\,(1+\lambda)}{\lambda-u^\star\,-\,0.62\,\lambda\,u^\star}\right)\label{allenformula}
\end{equation}
where 
\begin{equation}
\omega_{log}=\exp \left(\frac{2}{\lambda}\int_{0}^{\infty} d\omega \frac{\alpha^{2}F(\omega)}{\omega} \ln \omega \right) \label{eq13}
\end{equation}
represents the characteristic energy scale of phonons for pairing in the strong-coupling limit, while $f_1$, $f_2$ are semi-empirical correction factors, as defined in ~\cite{PhysRevB.12.905}.
The parameter $u^\star$ encodes the strength of the Coulomb interactions and it  is determined experimentally and tabulated in the literature for various materials; we will take it as an external input from tabulated literature data. That said, all the SC properties are determined by the shape of the spectral function $\alpha^2 F(\omega)$.
Finally, the SC gap can be estimated from:
\begin{equation}
    \frac{2\,\Delta}{k_B\,T_c}\,=\,3.53\,\left(1\,+\,12.5\,\left(\frac{T_c}{\omega_{log}}\right)^2\,\log \left(\frac{\omega_{log}}{2\,T_c}\right)\right).
\end{equation}

\section{Theoretical predictions}
\subsection{Linear in $\omega$ trend of $\alpha^{2}F(\omega)$ at low frequency}
Since, as will be shown below, the low-frequency behaviour of the Eliashberg function $\alpha^{2}F(\omega)$ is dominated by the two transverse modes (both with same frequency), we start by focusing on a single phonon branch in Eq.\eqref{aa}, to represent the transverse acoustic modes, with diffusion constant $D \equiv D_{T}$ and propagation speed $v \equiv v_{T}$.
\begin{figure}[h!]
    \centering
    \includegraphics[width=0.8\linewidth]{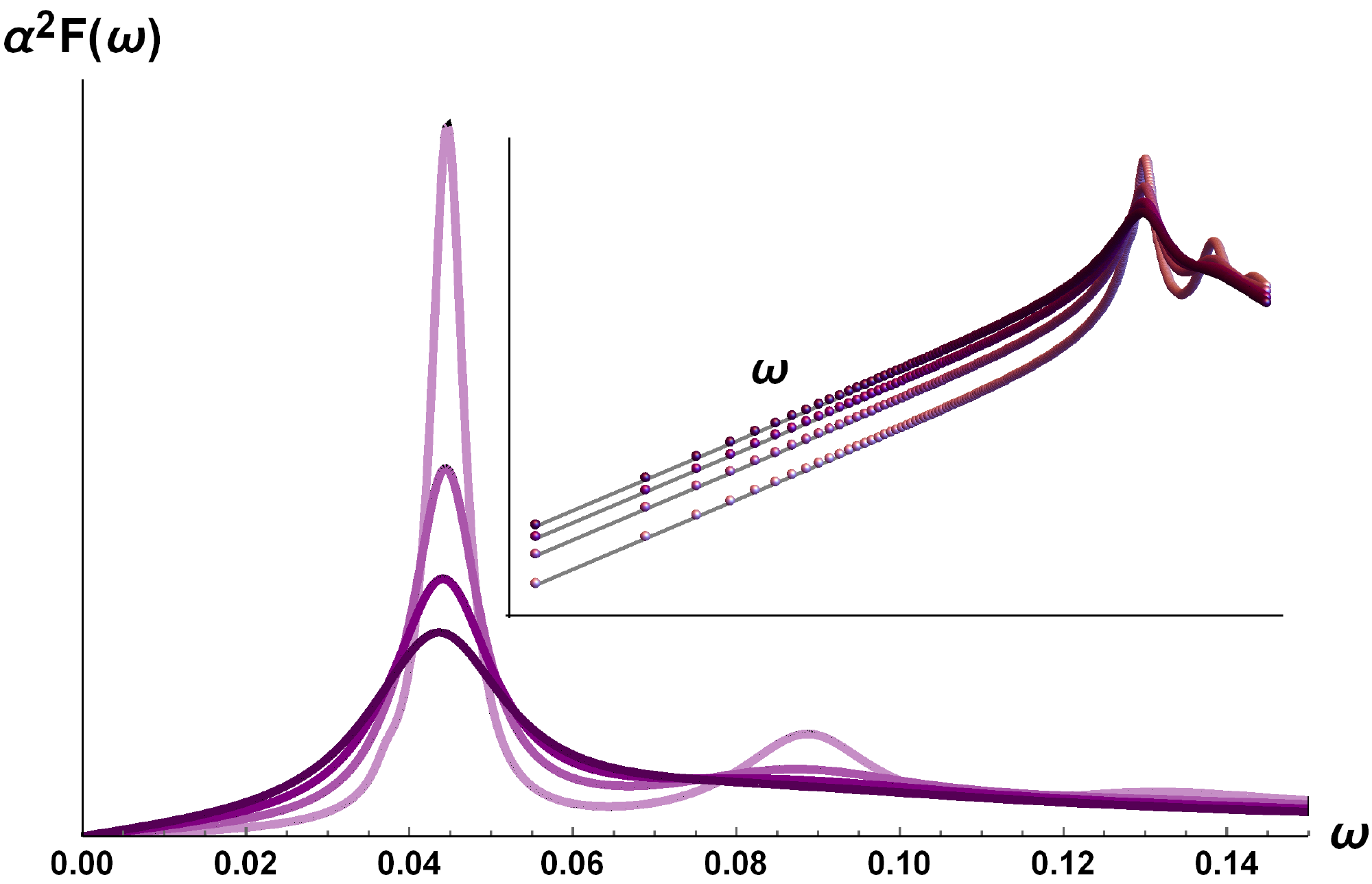}
    \caption{One-branch calculation of the electron-phonon spectral function $\alpha^2 F(\omega)$. The $y-$axes is shown in arbitrary units. The diffusion constant goes from $D=0.1$ (lighter line) to $D=2$ (darker line). The values of the other parameters are kept fixed. The inset shows the low frequency scaling $\sim \omega$ for the red line data. The scaling is independent of the value of the diffusion constant $D$.}
    \label{fig:1}
\end{figure}

In Fig.\ref{fig:1} we present theoretical predictions from the above model which show the linear in frequency behaviour of the Eliashberg function $\alpha^2 F(\omega)$ in the low $\omega$ limit. 
This behaviour was predicted by Bergmann~\cite{Bergmann1971} using a heuristic argument based on accounting for additional electron-phonon scattering processes due to lack of momentum conservation and of standard Umklapp conditions in disordered systems, and was confirmed experimentally in metal alloys~\cite{Keck1976}.

In our model, the linear behaviour in $\omega$ of the Eliashberg function at low $\omega$ comes directly from the form of the Lorentzian used to model the phonon spectral function, cfr. Eq.\eqref{eq4} and Eq.\eqref{eq9}. 
A similar result was derived, also using a Lorentzian for damped phonons, by Belitz~\cite{Belitz1987} focusing on metals with impurities where the damping originates from anharmonicity promoted by the defects. In our case, the effective damping $D$ is not originated by anharmonicity (which anyway at low $T$ is always low) but rather from the diffusive motion of vibrational excitations, hence by the diffusivity of the "diffusons" in a strong-disorder context.

\subsection{Increasing and non-monotonic behaviour of $\lambda$ with disorder}
\begin{figure}[h!]
    \centering
    \includegraphics[width=0.7\linewidth]{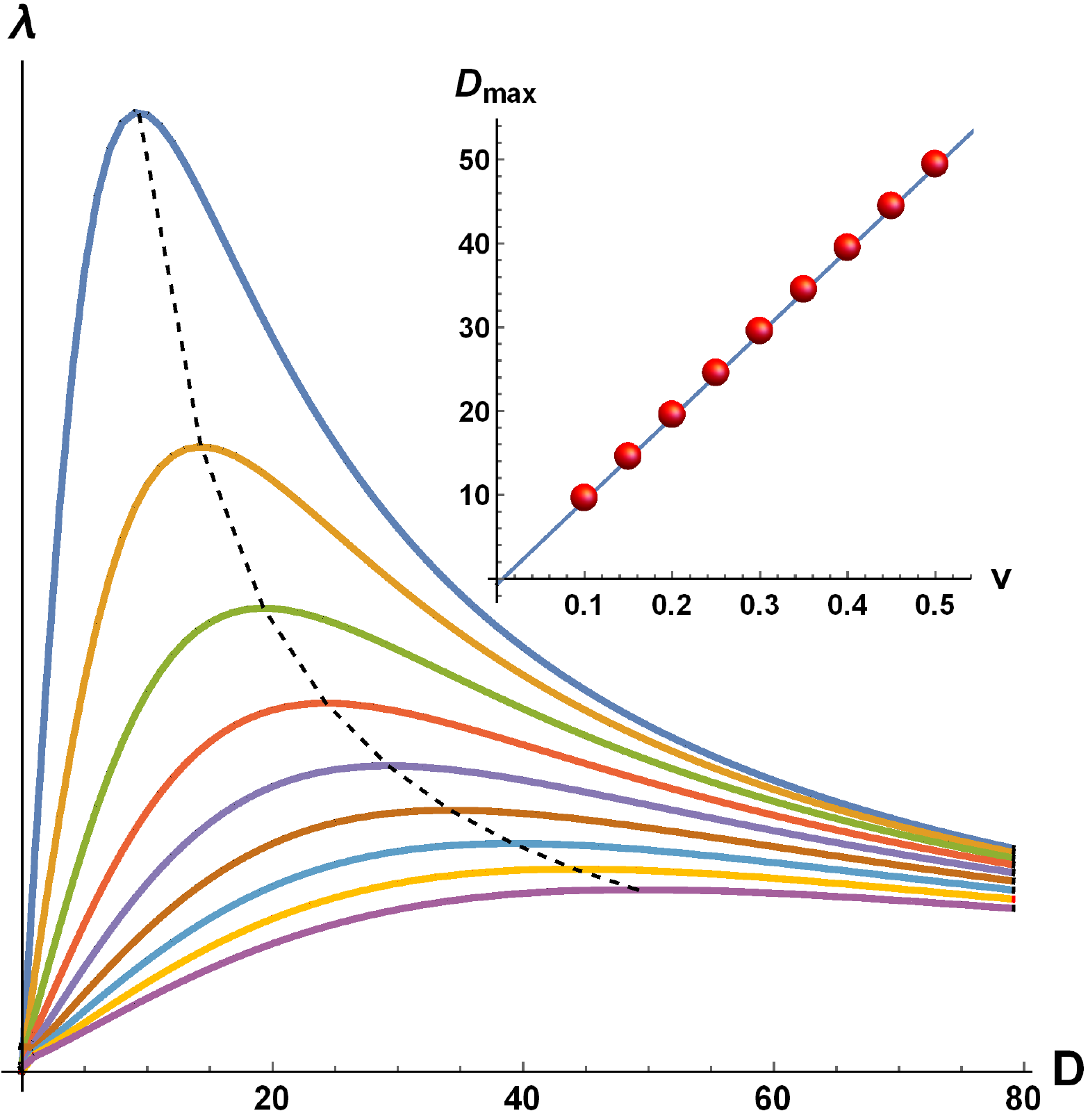}
    \caption{Electron-phonon coupling constant $\lambda$ as a function of the diffusivity $D$ of transverse vibrational excitations, for various values of the transverse speed of sound. The units for both axes are arbitrary. The inset shows how the position of the maximum varies as a function of the speed of sound; the curve is fitted very well by a linear relation (solid line). The relative amplitude of the peak between the line at large $v$ (bottom purple line) and the one at small $v$ (top blue line) where the peak is maximum, is about a factor $5$.}
    \label{fig:2}
\end{figure}

In Fig.\ref{fig:2} we plot the behaviour of the electron-phonon coupling parameter $\lambda$ evaluated using Eqs.\eqref{aa}-\eqref{eq11} as a function of the diffusivity $D$ of the vibrational excitation, which represents a parameter directly linked with structural disorder in the material. This is because upon increasing the disorder, the number of scattering events increases, thus leading to a larger value of $D$~\,\,\,\cite{Allen_diffusons,Parshin1,Parshin2,BaggioliPRR}.

A non monotonic behaviour of $\lambda$ as a function of the diffusion constant $D$ is shown in Fig.\ref{fig:2} for sufficiently low values of speed of sound $v$. The height of the peak decreases with the speed of sound $v$, and at the same time the position of the peak shifts to larger values of $D$. The position of the peak $D_{max}$ increases linearly with $v$, as shown in the inset of Fig.\ref{fig:2}. 
The non-monotonicity strongly depends on the speed of sound $v$, as shown in Fig.\ref{fig:4}, with the peak becoming less prominent (broader  and lower) and shifted to larger $D$ values as the speed of sound $v$ increases. 

The non-monotonic dependence of $\lambda$ upon the disorder parameter $D$ may be ascribed to the fact that increasing the disorder brings about two competing effects. On one hand, the Lorentzian vibrational peak becomes bigger (since $D$ sits in the numerator of the Lorentzian), which makes more phonon states accessible for pairing at low $\omega$ (again, in the absence of restrictions put by momentum conservation and Umklapp conditions, as already noted by Bergmann~\cite{Bergmann1971}). We note that the proliferation of low-frequency transverse excitations plays a crucial role in the $\lambda$ enhancement due to the $1 / \omega$ weight factor in the Eliashberg integral. 
On the other hand, however, upon increasing $D$ further, the Lorentzian becomes broader and eventually shallower (less peaked) due to the term $\sim D^{2}$ in the denominator of the Lorentzian. This can be seen as a manifestation of the fact that the lifetime of the vibrational excitation, $\tau \sim D^{-1}$, becomes too short to allow for efficient pairing (linewidth broadening with disorder).\\

The increasing part of the curve, may explain the increase of $T_{c}$ found (rarely) in certain experimental systems. However, one should also consider how the overall energy scale $\omega_{log}$ for strong-coupling superconductivity changes upon varying the disorder.

In order to illustrate the interplay between these opposing tendencies, in the next Sections we apply the above model to real data of amorphous metal and alloys.

\section{Effect of disorder on $T_c$: the case of $\text{Pb}$-based materials}
We consider the case of Pb as a model material for which experimental data are available, where the structural disorder of the sample can be varied continuously from the pure crystalline material limit (zero disorder) to the amorphous alloy (strong disorder)~\cite{Knorr1971}, by means of alloying. Furthermore, a mildly disordered state is obtained experimentally from a fine granular mixture of tiny crystallites, e.g. an example of a granular superconductor state~\cite{Knorr1971,Bergmann1976}.

\begin{figure}[h!]
    \centering
    \includegraphics[width=0.9 \linewidth]{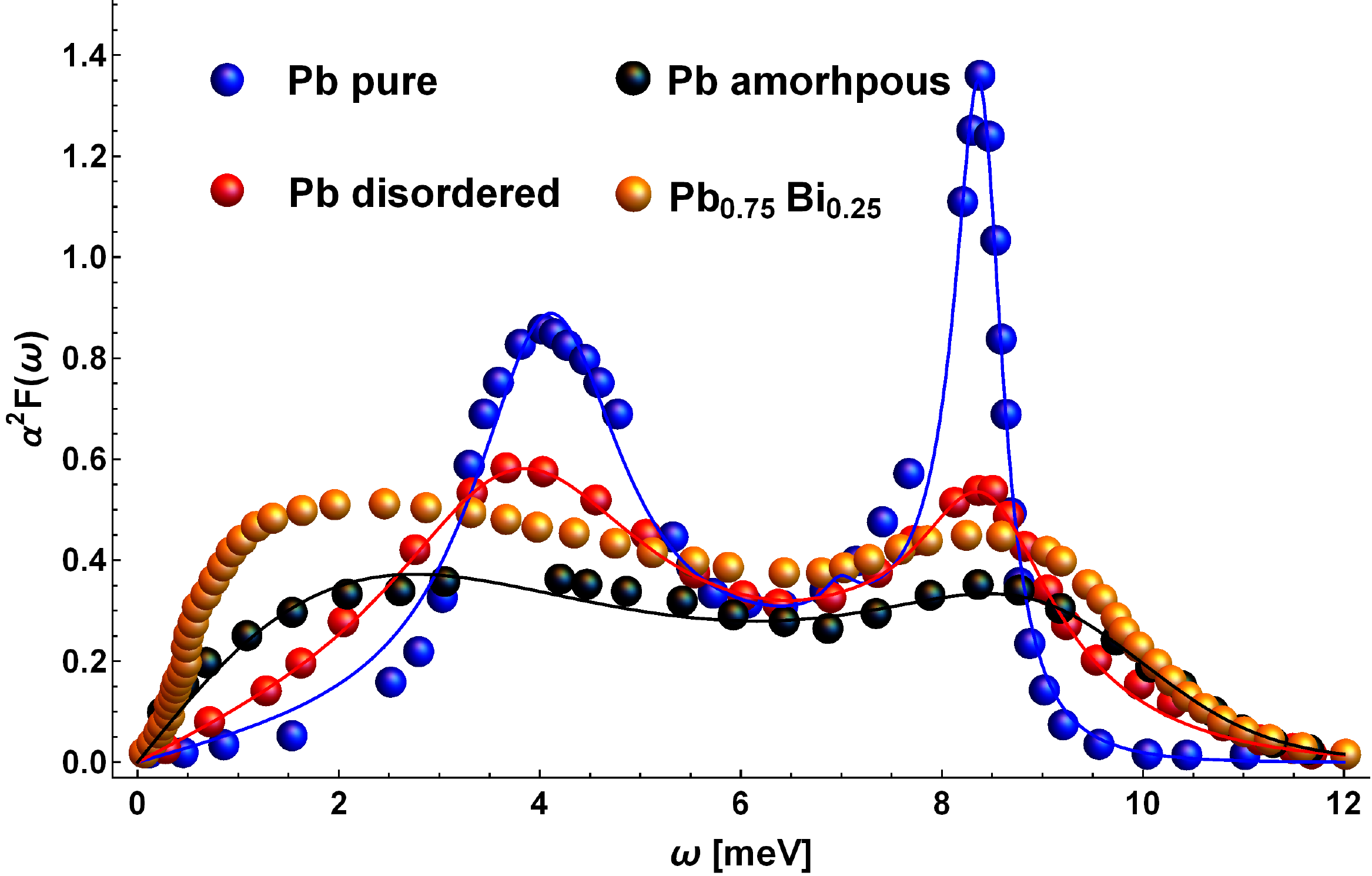}
    \caption{Theoretical model fittings of the electron-phonon spectral function for Pb-based materials. The data are taken from Ref.\cite{Knorr1971},and refer to pure crystalline Pb (blue), granular microcrystalline Pb (red), amorphous alloy $\text{Pb}_{0.9}\text{Cu}_{0.1}$ (black) and amorphous alloy $\text{Pb}_{0.75}\text{Bi}_{0.25}$ (orange), the latter taken from Ref.~\cite{Bergmann1973}.}
    \label{fig:3}
\end{figure}

\begin{figure}[h!]
    \centering
    \includegraphics[width=0.75\linewidth]{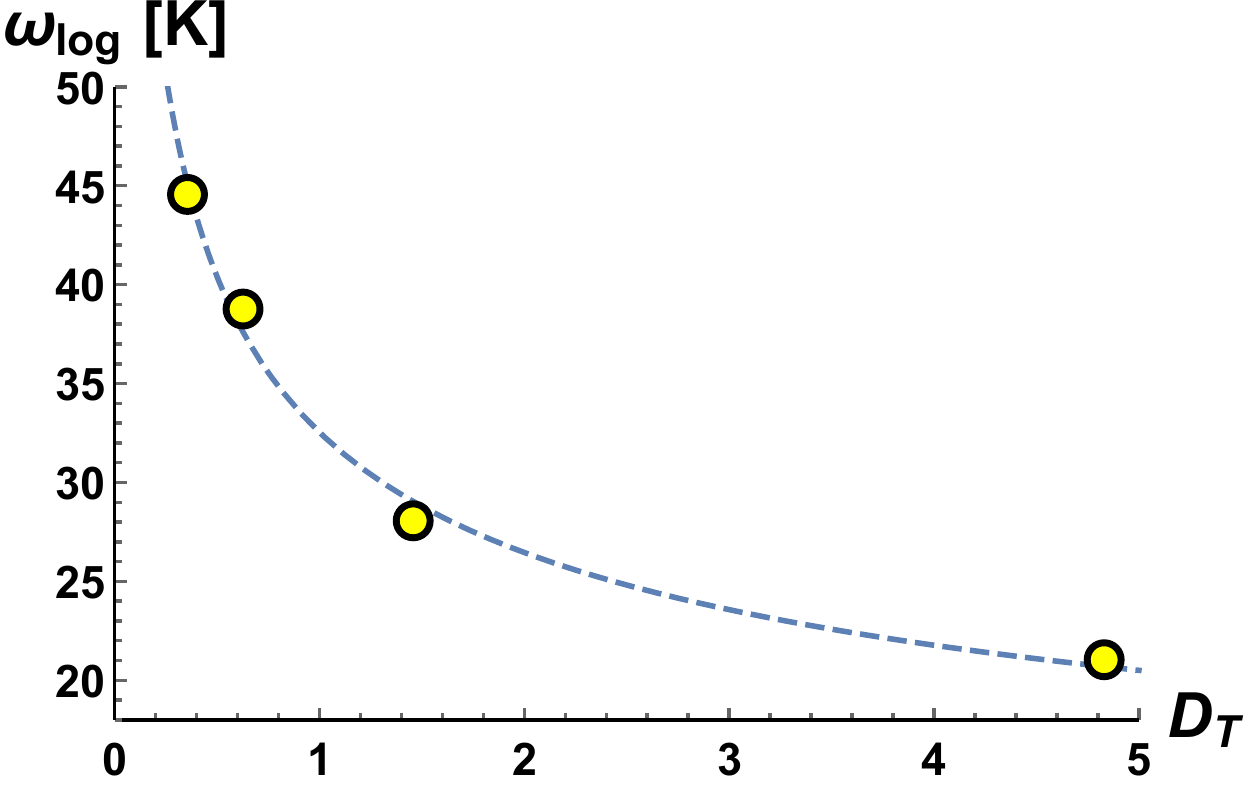}
    
    \vspace{0.2cm}
    
    \includegraphics[width=0.7\linewidth]{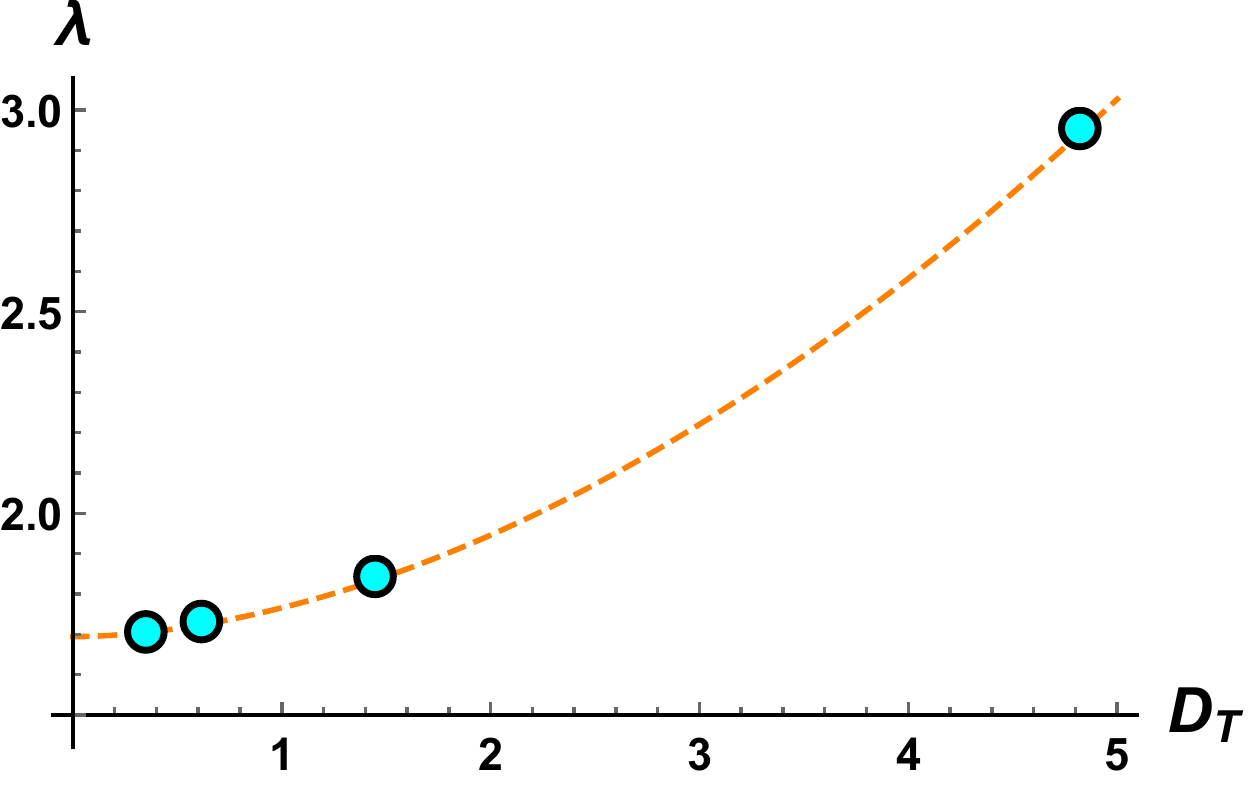}
    
    \vspace{0.2cm}
    
    \includegraphics[width=0.7\linewidth]{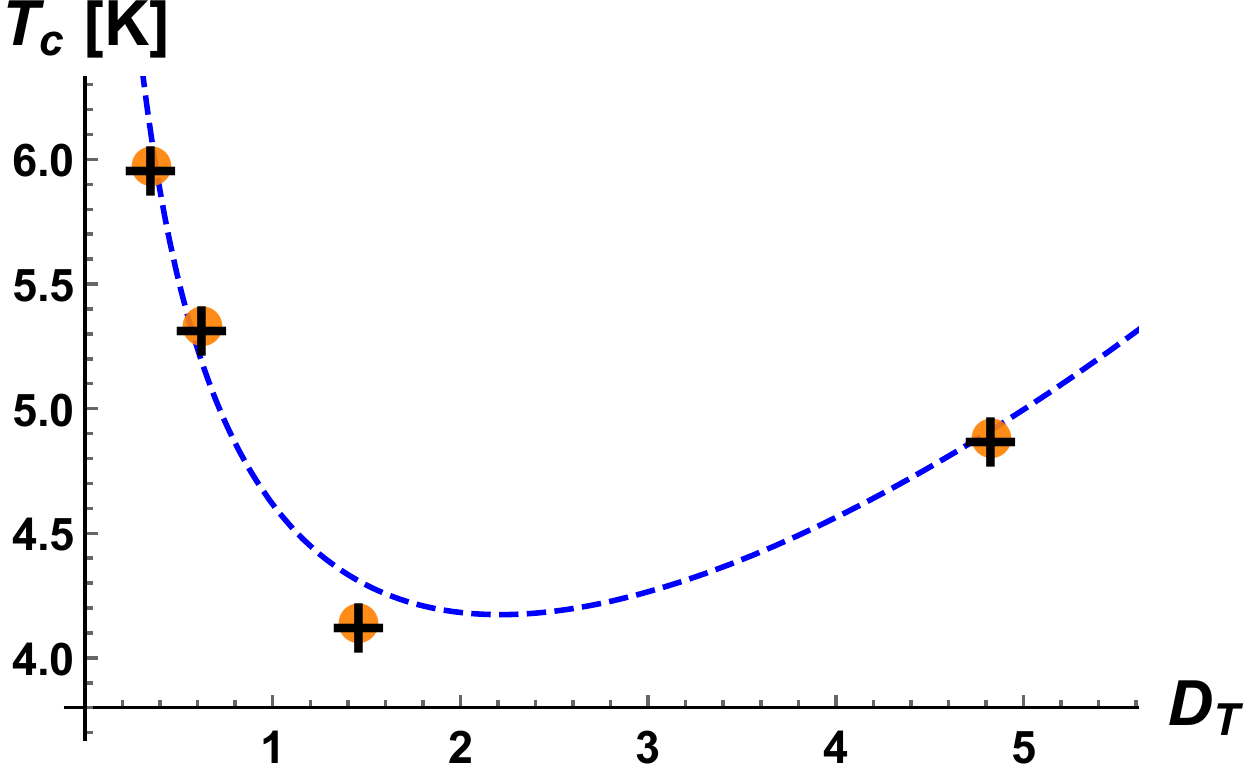}
    \caption{Behavior of characteristic physical parameters describing superconductivity for the four different Pb-based systems with varying disorder as a function of the transverse diffusivity parameter $D_{T}$. Symbols refer to the values of the three different systems in Ref.~\cite{Knorr1971} for pure Pb, granular and Ref.~\cite{Bergmann1973}, while dashed lines are empirical fitting trends. \textbf{Top panel}: the characteristic energy scale (prefactor in the Allen formula for $T_c$), $\omega_{log}$. \textbf{Middle panel}: the electron-phonon coupling constant $\lambda$ for the four systems. \textbf{Bottom panel}: the critical temperature for the onset of superconductivity, $T_c$ calculated from the Allen formula Eq. (12) in the text. For reference, the experimentally measured $T_{c}$ values are (in order of increasing $D_{T}$): $7.2$K (pure Pb), $7.19$K (granular microcrystalline), $6.5$K ($\text{Pb}_{0.9}\text{Cu}_{0.1}$), $6.9$ ($\text{Pb}_{0.75}\text{Bi}_{0.25}$). \color{black} These data are explicitly tabulated in \cite{data1,data2}, where also the original experimental references are quoted.\color{black}}
    \label{fig:4}
\end{figure}

As shown in Fig.\ref{fig:3}, these four states characterized by different degrees of disorder, can be fitted very nicely using our diffusons model, Eq.\eqref{aa} using $D_L$ and $D_T$ as fitting parameters (see Table \ref{tab1} for the values of the fitting parameters), while the ratio of $v_{L}/v_{T} \sim 1.8$ is fixed to typical values of isotropic amorphous materials~\cite{Parshin2}. In particular, the experimental Eliashberg functions features two distinct peaks, at low and high $\omega$, which correspond to the transverse (low frequency) and longitudinal (high frequency) vibrational excitations, respectively. 
It is evident that, upon increasing the alloying and hence the disorder, both peaks become much broader which corresponds to a significant increase of both $D_L$ and $D_T$, i.e. of the diffusivity of the vibrational excitations. The increase and the broadening, are comparatively much bigger for the transverse excitations which goes along with the boson peak effect in the VDOS~\cite{Schirmacher,baggioli2019unified}. This effect is important for the creation of new pairing states, hence for the enhancement of $\lambda$. 
At the same time, however, the softening of the transverse excitations causes a monotonic lowering of the overall energy scale $\omega_{log}$ upon increasing the disorder parameter $D \equiv D_T$, as displayed in Fig.\ref{fig:4} (top panel), obtained upon using the theoretical fitting with Eq.\eqref{aa} of the experimental data inside Eq.\eqref{eq13}. The trend of $\omega_{log}$ vs $D_{T}$ is well fitted by $\omega_{log} = 5.96 + 26.6 D_{T}^{-0.37}$, hence, importantly, it will eventually saturate to a constant (see below for the important consequences of this trend).

Using the theoretical fittings of the experimental data with Eq.\eqref{aa} inside Eq.\eqref{eq11}, now considering all three branches, one longitudinal and two transverse, we obtain the behaviour of the electron-phonon coupling constant $\lambda$ as a function of disorder, here quantified by $D_{T}$, displayed in Fig.\ref{fig:4} (central panel). Clearly, $\lambda$ presents a monotonic increasing trend as a function of $D_{T}$, which  is well fitted by a power-law function $\lambda = 1.69 + 0.07 D_{T}^{1.82}$, which  is the dashed line in the plot. Hence, disorder promotes the electron-phonon coupling in this system, as theoretically predicted by the model (Fig.\ref{fig:2}).
  
Using $\lambda$ and $\omega_{log}$ determined in this way inside the Allen formula Eq.\eqref{allenformula}, we obtain the behaviour of $T_c$ as a function of disorder in Fig.\ref{fig:4} (bottom panel). Because of the comparatively faster decrease of $\omega_{log}$ with increasing disorder, the initial trend is a monotonic decrease of $T_c$ upon increasing $D_{T}$. This estimated trend is in qualitative agreement with the reported experimental trend for $T_c$ for these systems, where $T_c \simeq 7.2$K for both the crystalline and the granular (mildly disordered) sample, while $T_c \simeq 6.5$K for the amorphous $\text{Pb}_{0.9}\text{Cu}_{0.1}$ alloy~\cite{Knorr1971}, thus presenting also a monotonic decrease with disorder.

However, upon increasing the alloying, which in turn increases the structural disorder further, the $\text{Pb}_{0.75}\text{Bi}_{0.25}$ alloy presents a higher $T_{c}$ compared to the less disordered $\text{Pb}_{0.9}\text{Cu}_{0.1}$. This fact suggests that, at sufficiently strong disorder, the $T_{c}$ becomes an increasing function of structural disorder (and of the degree of alloying, as in this case). 
This is explained with the fact that $\omega_{log}$ eventually saturates to a horizontal asymptote upon further increasing $D_{T}$, whereas $\lambda$ keeps increasing with $D_{T}$ in an unbound way. 
Using the empirical fitting functions presented above for $\lambda$ and $\omega_{log}$  inside the Allen formula we obtain the dashed line in Fig.\ref{fig:4} (bottom panel) where the upturn enhancement after the minimum clearly reflects the fact that $\omega_{log}$ eventually saturates, whereas $\lambda$ does not. 

The trend is indeed explained with the competition between $\omega_{log}$, which is a monotonic decreasing function of disorder due to the broadening of transverse excitations at low energy as seen above (which lowers the characteristic energy scale), and $\lambda$, which is a monotonically increasing function of disorder in the system, in the Allen formula Eq.\eqref{eq13}. It is clear from Fig.\ref{fig:4}(a) that there is a saturation in the decrease of $\omega_{log}$ with increasing disorder upon going from $\text{Pb}_{0.9}\text{Cu}_{0.1}$ to $\text{Pb}_{0.75}\text{Bi}_{0.25}$, and the respective values of $\omega_{log}$ are very comparable, while at the same time $\lambda$ keeps increasing with disorder.  
This is, once again, fully in agreement with experimental measurements \color{black}(tabulated in \cite{data1,data2})\color{black}, where $T_{c}\simeq 6.5$ for $\text{Pb}_{0.9}\text{Cu}_{0.1}$ while $T_{c}\simeq 6.9$ for $\text{Pb}_{0.75}\text{Bi}_{0.25}$.

This analysis suggests a possibly important principle for material design: while structural disorder promotes the enhancement of electron-phonon coupling hence of $\lambda$, at the same time the softening and broadening of the transverse excitations (related to the boson peak  phenomenon) should be compensated by a broadening of the longitudinal excitations towards higher frequency so as to saturate the decrease of $\omega_{log}$ with disorder. This is indeed the case in the above analysis where the "saturation" of $\omega_{log}$ upon going from $\text{Pb}_{0.9}\text{Cu}_{0.1}$ to $\text{Pb}_{0.75}\text{Bi}_{0.25}$ is controlled by the increase of $D_{L}$ which increases nearly by a factor two.
Our analysis of the Pb-based alloys shows that indeed upon increasing the alloying there is a creation of pairing states at high frequency promoted by the broadening of the longitudinal excitations which arrests the decrease of $\omega_{log}$ and thus inverts the decreasing trend of $T_{c}$ with disorder.


\begin{table}[h!]
    \centering
\begin{tabular}{ |c||c|c|c|c|  }
 \hline
 \multicolumn{5}{|c|}{Pb-based materials} \\
 \hline
 & pure & granular & $\text{Pb}_{0.9}\text{Cu}_{0.1}$ & $\text{Pb}_{0.75}\text{Bi}_{0.25}$ \\
 \hline
 $D_T$   & 0.358    & 0.582 &   1.735  & 4.90057\\
 $D_L$ &   0.116  & 0.348   & 0.366  & 0.55303\\
 \hline
\end{tabular}
\caption{Data of vibrational excitation diffusivities obtained from the fit in Fig.\ref{fig:3} using the theoretical model, Eq.\eqref{aa}.}
\label{tab1}
\end{table}

\section{Enhancement of $T_c$ induced by disorder: the case of Aluminum}
\begin{figure}[h!]
    \centering
    \includegraphics[width=0.9\linewidth]{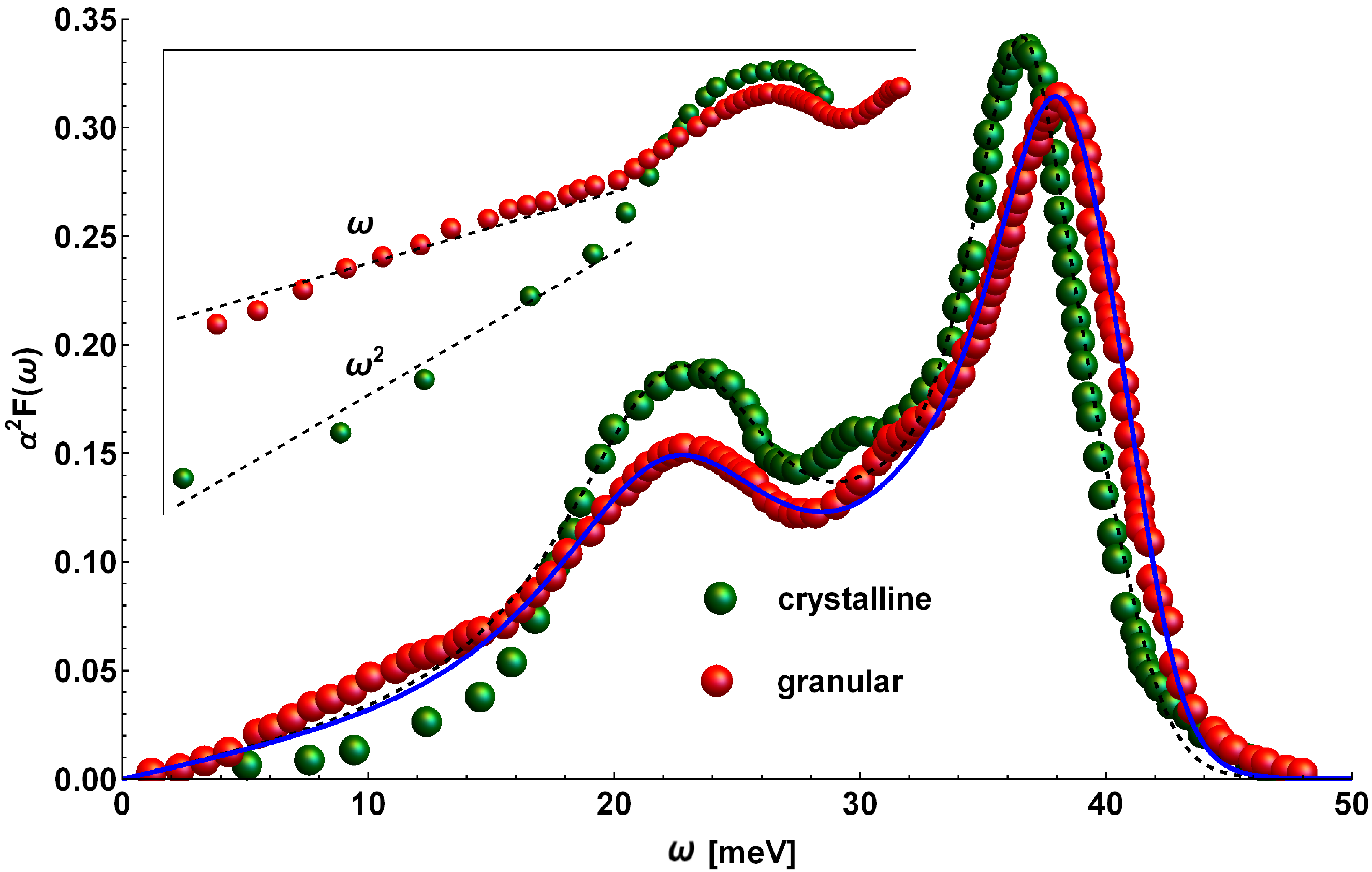}
    \caption{Theoretical modelling of Eliashberg function for Al-based materials. Symbols: experimental data of Eliashberg function $\alpha^{2}F(\omega)$ measured by electron tunnelling from Ref.~\cite{Dayan}. Continuous lines: model fitting for crystalline Al (solid line) and granular Al (dashed line). The inset show the different scaling in the low-energy limit for crystalline and granular Al. }
    \label{fig:5}
\end{figure}

A  famous example of a system where disorder leads to an enhancement of $T_{c}$ is represented by Al. Experimental data of the Eliashberg function $\alpha^{2}F(\omega)$ were obtained by Dayan using electron tunnelling measurements on both crystalline Al and granular Al~\cite{Dayan}. 
The experimental data have been fitted with our theoretical model, Eqs.\eqref{aa}-\eqref{eq10} and the comparison is shown in Fig.\ref{fig:5}. 
Thanks to the cubic structure of Al, also here the main peaks refer to acoustic transverse excitations (the low-energy peak) and to longitudinal excitations (the high-energy peak), while optical modes provide negligible features in comparison. It is seen that the theoretical model provides an excellent fitting for the granular Al  where it perfectly catches the $\sim \omega$ trend at low $\omega$ typical of amorphous superconductors. The fitting is less accurate for the crystalline sample at low $\omega$ where instead $\alpha^{2}F(\omega) \sim \omega^{2}$ at low energy. This is due to the fact that in the crystalling sample there is a clean Debye $\sim \omega^{2}$ scaling which  extends over a broad energy range, whereas in the granular sample the structural disorder induces a deviation from Debye's law with an excess of low-energy modes, closely related to the boson peak phenomenon~\cite{Schirmacher,Milkus}.
This excess of vibrational modes thus clearly gives an excess of pairing states which leads to an enhancement of $\lambda$ in the granular Al with respect to crystal Al, due to the additional $1/\omega$ weight in the definition of $\lambda$, see Table \ref{tab2} for the values of $\lambda$. 

On the other hand, however, the excess of low-energy modes also causes a decrease of $\omega_{log}$ upon going from crystalline to granular, see Table \ref{tab2} for the respective values.
If the decrease $\omega_{log}$ however is not too big to cancel the enhancement due to increased $\lambda$ in the Allen formula, then an overall enhancement of $T_{c}$ is possible.
This is indeed the case of Al, and we can explain this again with the fact that the high-energy longitudinal peak in Fig.\ref{fig:5} gets broader and moves to higher energy for the granular sample, hence $D_{L}$ increases quite significantly from $1.15$  for crystalline to $1.6$ for granular Al (Table \ref{tab2}). 

This analysis is fully consistent with the conclusions drawn in the analysis of Pb-based systems  and fully confirms that an enhancement of $T_{c}$ in amorphous systems is achievable whenever both the transverse $D_{T}$
and the longitudinal $D_{L}$ phonon diffusivities increase upon going from crystalline to disordered. While the increase of $D_{T}$, related to the boson peak in the VDOS, promotes the increase of $\lambda$, the increase of $D_{L}$ puts a bound on the decrease of $\omega_{log}$, thus leading to an overall increase in $T_{c}$.

\begin{table}[h!]
    \centering
\begin{tabular}{ |c||c|c|  }
 \hline
 \multicolumn{3}{|c|}{Aluminium} \\
 \hline
 & crystalline & granular  \\
 \hline
 $D_T$ (fit)   & 2.2    & 2.4 \\
 $D_L$ (fit) &   1.15 & 1.6\\
 $\lambda$ (interpolation)   & 0.358    & 0.410 \\
 $\lambda$ (fit)   & 0.376    & 0.430 \\
 $\omega_{log}$ (interpolation)   & 25.2 meV  & 20.6 meV \\
 $\omega_{log}$ (fit)   & 23.2 meV    & 20.4 meV \\
 $T_c$ (interpolation)   & 0.60 K  &  1.71 K \\
 $T_c$ (fit)   &  0.76 K  & 2.11 K \\
 \hline
\end{tabular}
\caption{Values of parameter found in the analysis of crystalline and granular Al using our model. The values of $D_L$ and $D_T$ are obtained by fitting the experimental $\alpha^{2}F(\omega)$ from Ref.~\cite{Dayan} to our model. The values of $\lambda$, $\omega_{log}$ and $T_{c}$ are either obtained using the fitted function $\alpha^{2}F(\omega)$ from our model, "(fit)", or from a spline interpolation to the experimental data , "(interpolation)".}
\label{tab2}
\end{table}

\section{Summary}
We developed a theoretical modelling framework for the superconductivity of amorphous materials based on the effective concept of \textit{diffusons}~\cite{Allen_diffusons,BaggioliPRR} to quantitatively describe the underlying vibrational excitations which are significantly affected by disorder-induced scattering phenomena. 
The model leads to an analytical form for the Eliashberg function $\alpha^{2}F(\omega)$ in terms of the transverse and longitudinal speeds of sound $v_{T},v_{L}$ and of the diffusivities $D_{T}$ and $D_{L}$ of transverse and longitudinal vibrational excitations (the latter are also related to sound absorption coefficients~\cite{Belitz1987}). In particular, $D_{T}$ controls the broadening of the transverse excitations which is very significant upon increasing the disorder. Hence, being this parameter particularly sensitive to structural disorder, it provides a natural way to quantify the extent of structural disorder in a given material.

The model provides a number of predictions: (i) it recovers the linear-in-$\omega$ trend of $\alpha^{2}F(\omega)$ at low $\omega$ ~\cite{Bergmann1971,Bergmann1976,Belitz1987}; (ii) it predicts a non-monotonic behaviour of the electron-phonon coupling $\lambda$ with a maximum as a function of disorder, which turns into a monotonic increasing trend upon increasing the speed of sound; (iii) it predicts a linear increasing dependence of the maximum in $\lambda$ vs disorder as a function of the speed of sound.

In particular, the increase of $\lambda$ with disorder is controlled by the proliferation of low-frequency transverse excitations which contribute decisively inside the Eliashberg integral due to the $1/\omega$ weight in the integral. Even though $\lambda$ is seen to increase with disorder, either monotonically or up to some maximum, the net effect of disorder on $T_{c}$ can induce a monotonic decrease of $T_{c}$ with disorder because the characteristic energy scale $\omega_{log}$ decreases significantly (and often comparatively faster) due to the proliferation of low energy transverse excitations upon increasing the disorder.

The model can be used to fit experimental data of $\alpha^{2}F(\omega)$ in an attempt to rationalize the effect of structural disorder on $T_{c}$ across a broad spectrum of materials. From the fitting, the values of vibrational line-widths $D_{T}$ and $D_{L}$ can be extracted as an output, which can be used to effectively quantify and parameterize the extent of disorder in specific materials. 

As an example to illustrate the usefulness of the approach in rationalizing a scattered amount of experimental data on amorphous materials, we applied the theoretical model to experimental data of Pb-based materials. In these systems, the disorder can be varied all the way from the pure crystalline Pb to alloys with increasing disorder. Since crystalline Pb is already a good superconductor owing to its close-packed FCC structure~\cite{Bergmann1976,Buzea_2004}, introducing disorder leads to a decrease of the $T_{c}$ even though $\lambda$ increases monotonically with increasing disorder. The net decrease is controlled by the decrease of the characteristic energy scale $\omega_{log}$ with disorder, due to the proliferation of low-energy vibrational excitations which cause the decrease of $\omega_{log}$.
However, upon increasing the alloying further, the decreasing trend of $\omega_{log}$ appears to saturate and no longer wins over the enhancement of $\lambda$ due to disorder. This leads to an inversion of the trend and $T_{c}$ is shown to increase upon increasing the disorder. The predicted trend for $T_{c}$ is in qualitative agreement with experimental measurements, which also feature this trend inversion. 

All in all, the proposed model could be useful to rationalize a large amount of scattered experimental evidence about the effect of structural disorder on superconductivity, including recent developments~\cite{Leroux2019,Zhao2019} and as a means to provide chemical-design principles for optimizing the $T_{c}$ of amorphous superconductors.\\

\section*{Acknowledgements}
We thank Zvi Ovadyahu for bringing Ref.~\cite{Dayan} to our attention. Useful discussions and input by Boris Shapiro and Miguel Ramos are gratefully acknowledged. 
M.B. acknowledges the support of the Spanish MINECO’s
“Centro de Excelencia Severo Ochoa” Programme under
grant SEV-2012-0249. CS is supported by DOE grant number DE-FG02-05ER46236.
 
\appendix

\bibliographystyle{apsrev4-1}

\bibliography{amorphous}

\end{document}